\documentclass[10pt,twocolumn,superscriptaddress,longbibliography,aps,prb,preprintnumbers]{revtex4-2}

\usepackage[normalem]{ulem}
\usepackage{graphicx}
\usepackage{bm}
\usepackage{color}
\usepackage{epstopdf}
\usepackage{amsmath}
\usepackage{amssymb}
\usepackage{epstopdf}
\usepackage{lipsum}

\usepackage{gensymb}

\usepackage{multirow}

\usepackage{xfrac}

\usepackage{scrextend}

\usepackage[urlcolor=blue,colorlinks=true,citecolor=blue,linkcolor=blue,pdfstartview={FitH},bookmarks=false]{hyperref}

\graphicspath{{fig/}{./fig/}{.}}

\begin{document}

\title{
Electronic topological modes in hexagonal NbSi$_{2}$ with a double helix
}

\author{Issam Mahraj}
\email[e-mail: ]{issam.mahraj@ifj.edu.pl}
\affiliation{Institute of Nuclear Physics, Polish Academy of Sciences, W. E. Radzikowskiego 152, PL-31342 Krak\'{o}w, Poland}

\author{Andrzej Ptok}
\email[e-mail: ]{aptok@mmj.pl}
\affiliation{Institute of Nuclear Physics, Polish Academy of Sciences, W. E. Radzikowskiego 152, PL-31342 Krak\'{o}w, Poland}

\date{\today}

\begin{abstract}
Chiral systems can host nontrivial topological states, including characteristic edge modes and Fermi arcs.
Here, we investigate the electronic properties of hexagonal NbSi$_2$ with chiral symmetry using first-principles calculations for both bulk and slab geometries.
We find that spin--orbit coupling (SOC) is relatively weak.
Notably, the topological edge state originates from structural chirality rather than the SOC.
From the Berry curvature analysis, we confirm the existence of Weyl points, some of which carry a Chern number of $\pm 2$.
Furthermore, we demonstrate the existence of chiral edge modes with extended Fermi arcs around high-symmetry points.
Finally, we investigate the real-space localization using spatially resolved and site-dependent density of states (DOS).
We demonstrate that the surface states can be exclusively associated with the chiral Si chains.
\end{abstract}

\maketitle

%%%%%%%%%%%%%%%%%%%%%
%%%%%%%%%%%%%%%%%%%%%
%%%%%%%%%%%%%%%%%%%%%

\section{Introduction}

Chirality, a property of objects that cannot be superimposed on their mirror images, plays an important role in determining their physical properties~\cite{barron.12,fecher.kubler.22,cheong.xu.22,bousquet.fava.25}.
As a symmetry property of a system, chirality can lead to the emergence of topological states.
This has been reported in several compounds, for both electronic~\cite{chang.xu.17,tang.zhou.17,rao.li.19,li.xu.19,
sanchez.belopolski.19,yuan.zhou.19,schroter.pei.19,
takane.wang.19,yao.manna.20,sessi.fan.20,dai.li.22,
dutta.ghosh.22,cochran.belopolski.23,mahraj.ptok.24} and phononic~\cite{zhang.song.18,miao.zhang.18,li.zhang.21,
mahraj.ptok.24,mahraj.ptok.25} systems.

The existence of chiral chains can give rise to topological surface states, including Fermi arcs connecting the surface projections of Weyl nodes.
In this context, the most well-known examples are monosilicides and monogermanides~\cite{chang.xu.17,tang.zhou.17,rao.li.19,li.xu.19,
sanchez.belopolski.19,yuan.zhou.19,takane.wang.19,dai.li.22,
dutta.ghosh.22,cochran.belopolski.23}, crystallized in the cubic $P2_13$ space group (SG).
For these systems, the electronic band structure contains two double Weyl points --- at $\Gamma$ and R points.
In the absence of the spin--orbit coupling (SOC), the electronic band structure hosts spin--1-like dispersion at $\Gamma$ point and doubly degenerate Dirac cones at R point.
Furthermore, extended Fermi arcs can connect the surface projections of the $\Gamma$ and R points.
The SOC leads to the lifting of these degeneracies and to a more complex band structure~\cite{mahraj.ptok.24}.

Here, we focus on the electronic properties of chiral hexagonal disilicide NbSi$_2$ with $P6_222$ chiral symmetry.
Its crystal structure contains double helix of Si atoms, which affect the electronic band structure.
For example, NbSi$_2$ exhibits a chirality-induced spin-selectivity effect~\cite{shiota.inui.21,shishido.sakai.21} --- an effect in which spin-polarized current is generated during electronic transport.
Moreover, the vibrational modes can also affect the transport properties because of the crystal symmetry~\cite{garcia.nenno.21}.
The chiral crystal structure can also give rise to topological surface states.
The presence of screw rotational symmetry can enforce nontrivial electronic band topology and promote the emergence of Weyl nodes.~\cite{tsirkin.souza.17}.
Recently, angle-resolved photoelectron spectroscopy (ARPES) measurements were reported~\cite{zhang.shishidou.23}.
However, the existence of chiral edge modes has not been discussed in detail so far.
In this paper, we investigate the electronic structure of NbSi$_2$ in both bulk and slab geometries to address this question.

The remainder of this paper is organized as follows.
Section~\ref{sec.method} describes the computational methods.
The electronic bulk properties are discussed in Sec.~\ref{sec.bulk}, while the surface properties are addressed in Sec.~\ref{sec.surf}.
A summary is given in Sec.~\ref{sec.sum}.

%%%%%%%%%%%%%%%%%%%%%
%%%%%%%%%%%%%%%%%%%%%
%%%%%%%%%%%%%%%%%%%%%

\begin{figure}[!b]
\includegraphics[width=0.8\linewidth]{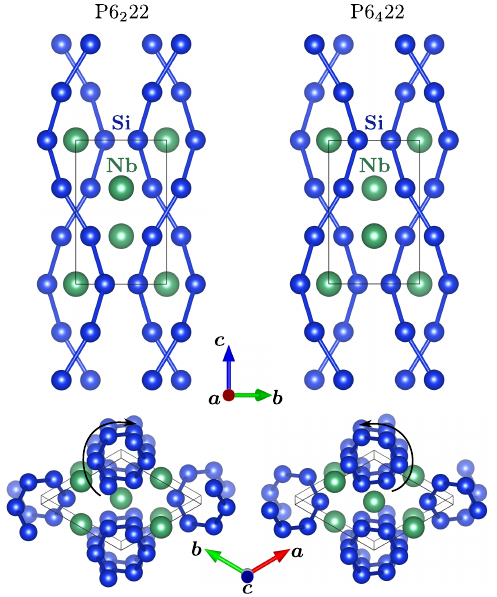}
\caption{
Crystal structure of NbSi$_{2}$ in two enantiomorphic chiral symmetries: SG No.~180, $P6_222$ and SG No.~181, $P6_422$ (left and right panels, respectively). 
The top and bottom panels show front and top views, respectively. Both structures possess chiral Si chains with opposite chirality, as indicated by the arrows in the bottom panels.
\label{fig.crys}}
\end{figure}

\section{Computational details}
\label{sec.method}

The first-principles (DFT) calculations were performed using the projector augmented-wave (PAW) potentials~\cite{blochl.94} implemented in the Vienna Ab initio Simulation Package ({\sc Vasp}) code~\cite{kresse.hafner.94,kresse.furthmuller.96,kresse.joubert.99}.
The calculations were performed within the generalized gradient approximation (GGA) using the Perdew–Burke–Ernzerhof (PBE) parameterization~\cite{perdew.burke.96}.
The plane-wave cutoff energy was set to $400$~eV.
The electronic configurations used in the calculations were $4p^{6}4d^{4}5s^{1}$ for Nb and $3s^{2}3p^{2}$ for Si.
The structural parameters, including lattice constants and atomic positions, were optimized in the conventional unit cell using a $12 \times 12 \times 8$ ${\bm k}$-point mesh within the Monkhorst--Pack scheme~\cite{monkhorst.pack.76}.
The structural optimization was considered converged when the total-energy differences were below $10^{-6}$~eV for ionic degrees of freedom and $10^{-8}$~eV for electronic degrees of freedom.

Similarly, slab calculations were performed using {\sc VASP}. 
The slab geometry consisted of 18 formula units (i.e., 54 atoms), and a $12 \times 12 \times 1$ $\Gamma$-centered ${\bm k}$-point mesh was employed. 
The real-space characteristics of the resulting states were analyzed using in-house software.

The tight-binding model in a Wannier-orbital basis~\cite{marzari.mostofi.12,marzari.vanderbilt.97,souza.marzari.01} constructed from the exact DFT electronic band structure calculations using the {\sc Wannier90} software~\cite{mostofi.yates.08,mostofi.yates.14,pizzi.vitale.20}. 
For these calculations, a $\Gamma$-centered $18 \times 18 \times 4$ ${\bm k}$-point grid was used.
As an initial basis, we considered the $s$, $p$, and $d$ atomic orbitals of Nb atoms and the $s$ and $p$ atomic orbitals of Si atoms, yielding a tight-binding model comprising 51 Wannier orbitals (corresponding to 102 bands). 
The resulting model was symmetrized to restore the crystal symmetries and subsequently employed to investigate the topological properties of the system. 
The surface Green's function for the semi-infinite geometry~\cite{sancho.sancho.85} was calculated using {\sc WannierTools}~\cite{wu.zhang.18}.

%%%%%%%%%%%%%%%%%%%%%
%%%%%%%%%%%%%%%%%%%%%
%%%%%%%%%%%%%%%%%%%%%

\section{Bulk properties}
\label{sec.bulk}

\subsection{Crystal structure and system symmetries}

NbSi$_2$ crystal can crystallize in one of two {\it enantiomorphic} structures: a right-handed $P6_222$ (SG No.~180) or a left-handed $P6_422$ (SG No.~181) --- as shown in Fig.~\ref{fig.crys}.
Both structures are dynamically stable; i.e., there are no imaginary phonon modes in the phonon spectrum~\cite{mahraj.ptok.25}.
The ground-state energies are identical for both symmetries. 
Experimentally, the crystal handedness can be controlled during crystal growth using the laser-diode-heated floating-zone method~\cite{kousaka.sayo.23}.
However, the $P6_222$ structure is more preferred in nature~\cite{sakamoto.fujii.05}.

The optimized lattice parameters are $a = 4.8177$~\AA\ and $c = 6.6233$~\AA, which are in good agreement with the experimental values of $a = 4.819$~\AA\ and $c = 6.592$~\AA~\cite{kubiak.horyn.72}.
The primitive unit cell contains nine atoms, i.e., three formula units.
The atomic positions are highly symmetric: Nb occupying the $3c$ Wyckoff position $(\sfrac{1}{2},0,0)$, while Si occupying the $6i$ Wyckoff position $(x_\text{Si}, 2x_\text{Si}, 0)$, with $x_\text{Si} = 0.1591$.
In this crystal structure, the distance between Si atoms within a single chain is approximately $2.58$~\AA, while the bond angle (i.e., the angle between two neighboring bonds) is approximately $150.14^{\circ}$. Similarly, the in-plane distance between Si atoms belonging to different chains of the same double helix is $2.65$~\AA, while the corresponding distance between two different double helices is $2.85$~\AA.
These comparable interatomic distances suggest strong bonding between Si atoms within the same double helix, as well as between neighboring double helices.

Both crystal structures contain double helix of Si atoms along the $c$ axis, with opposite chiralities.
Such chains are analogous to the structure of deoxyribonucleic acid (DNA), with the Si chains winding around each other in a helical arrangement (Fig.~\ref{fig.crys}).
Within the helix, pairs of atoms lie in the same plane, with each plane rotated by $\pm 60^{\circ}$ relative to the neighboring plane, where the sign depends on the crystal chirality.

The crystal structure exhibits several important symmetries. 
The most prominent is a sixfold screw symmetry, $6_2$, along the $c$ axis, corresponding to a rotation of $\pm 60^\circ$ combined with a fractional translation of $c/3$ along the same direction. 
In addition, the structure possesses twofold rotational and screw axes, as well as twofold rotation axes perpendicular to the $c$ axis. 
Importantly, the crystal lacks both mirror planes and inversion centers. 
The absence of inversion symmetry allows for the emergence of Weyl nodes, while the nonsymmorphic symmetries play an important role in protecting the topological features of the electronic structure~\cite{tsirkin.souza.17,gonzalez.tuiran.20}.

\begin{figure*}[!ht]
\includegraphics[width=\linewidth]{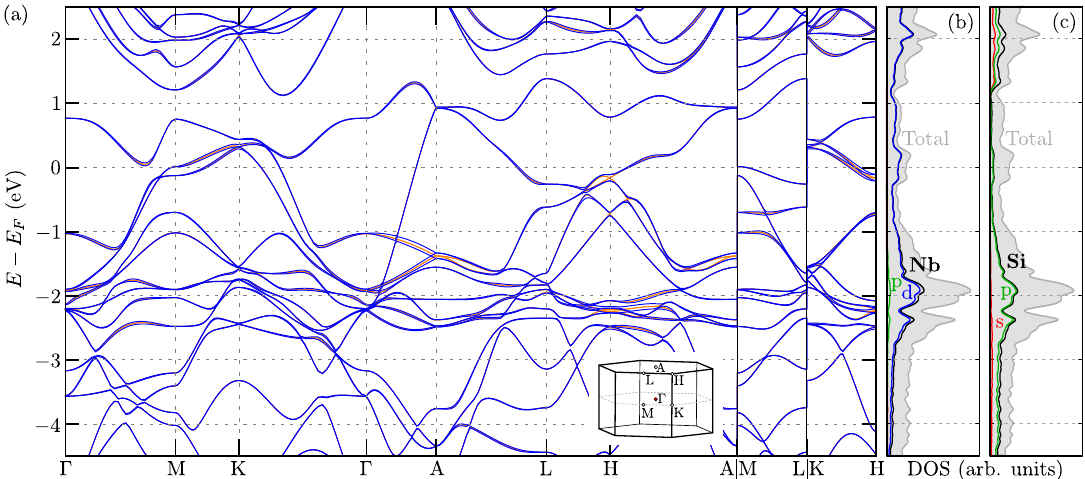}
\caption{
(a) The electronic band structure of NbSi$_{2}$ with $P6_222$ symmetry along high-symmetry directions.
The inset shows the corresponding Brillouin zone (BZ) and its characteristic high-symmetry points.
Results without and with the SOC are shown by orange and blue lines, respectively.
The orbital-projected partial density of states (DOS) for (a) Nb and (b) Si atoms. 
The total DOS of NbSi$_2$ is shown as a grey background.
\label{fig.el_band}}
\end{figure*}

\subsection{Electronic band structure}

The bulk electronic band structure and DOS are shown in Fig.~\ref{fig.el_band}.
The obtained band structure [Fig.~\ref{fig.el_band}(a)] is in qualitative agreement with earlier results from DFT calculations based on the local density approximation (LDA)~\cite{antonov.yavorsky.96}. 
However, the PBE-GGA parametrization leads to more subtle dispersion features.
Nevertheless, the orbital contributions remain largely unchanged and are dominated by Nb $d$ and Si $p$ states [see Fig.~\ref{fig.el_band}(b) and~\ref{fig.el_band}(c)]. 
The states around $-2$~eV, where the DOS is relatively high, can be characterized as a bonding combination of Nb $d$ and Si $p$ states. 
Similarly, an antibonding combination can be identified at energies above $2$~eV (not shown).
This DOS structure is consistent with experimental observations based on X-ray photoemission spectroscopy (XPS) and bremsstrahlung isochromat spectroscopy (BIS)~\cite{speier.kumar.89}. 
More recently, ARPES measurements have also revealed a group of electronic bands with substantial spectral weight around this binding energy~\cite{zhang.shishidou.23}.

Fig.~\ref{fig.el_band}(a) shows the electronic band structure in the absence and presence of the SOC.
The SOC leads to a surprisingly small band splitting (cf.~orange and blue lines).
Indeed, the SOC-induced splitting is appreciable only for the Nb $p$ orbitals, with an on-site SOC strength of approximately $15$~meV.
However, Nb $p$ states are located well below the Fermi level, at approximately $-30$~eV.
The SOC lifts the degeneracy of the Nb $4p$ states, splitting them into the $4p_{1/2}$ and $4p_{3/2}$ levels, with an energy separation of approximately 2~eV.
In contrast, the on-site SOC strengths for states near the Fermi level, which predominantly originate from Nb $d$ and Si $p$ orbitals, are smaller than 1.5~meV. 
Nevertheless, these contributions give rise to average band splittings of approximately 20--27~meV for bands crossing the Fermi level.

\begin{figure}[!b]
\includegraphics[width=\linewidth]{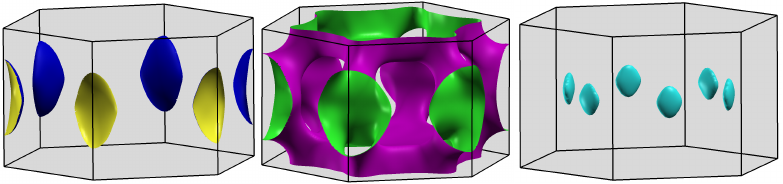}
\caption{
The Fermi-surface pockets of NbSi$_{2}$ with $P6_2 22$ symmetry in the absence of the SOC.
The uniform coloring distinguishes the inner and outer surfaces of each individual Fermi pocket.
\label{fig.fs}}
\end{figure}

In the absence of the SOC, the Fermi surface comprises three pockets (Fig.~\ref{fig.fs}).
Including the SOC doubles the number of Fermi pockets owing to the lifting of the band double degeneracy.
The bulk Fermi surface has also been experimentally investigated using de Haas--van Alphen (dHvA) measurements~\cite{onuki.nakamura.14}.
Interestingly, each observed dHvA branch was split into two separate branches by the SOC.
This feature was also observed in other transition-metal disilicides $T$Si$_2$ ($T=$ V, Nb, and Ta)~\cite{onuki.nakamura.14,tsutsumi.miura.13}.
For NbSi$_{2}$, the experimentally reported splitting energies are $209$, $244$, and $334$~K~\cite{onuki.nakamura.14}, corresponding to approximately $18$, $21$, and $28$~meV, respectively. These experimental values are in good agreement with the band splittings obtained from our calculations.
We also note that the bulk electronic band structures are very similar across the $T$Si$_2$ family ($T=$ V, Nb, and Ta)~\cite{tsutsumi.miura.13,roy.cerasoli.22}.
The primary difference is the increasing SOC strength with increasing atomic mass, from V to Ta.

\subsection{Bulk topological properties -- Weyl nodes}

In the presence of both time-reversal ($\mathcal{T}$) and inversion ($\mathcal{I}$) symmetries, all bands are double degenerated throughout the BZ.
Breaking of either $\mathcal{T}$ or $\mathcal{I}$ symmetries lead to emerge of Weyl nodes~\cite{armitage.mele.18}, which makes a touching between only two bands possible.
In NbSi$_{2}$, the $\mathcal{T}$ symmetry is preserved due to the absence of magnetism, while system is characterized by the lacks $\mathcal{I}$ symmetry.
Beyond, the existence of the screw rotational symmetry enforced the existence of the Weyl nodes~\cite{tsirkin.souza.17}, while the topological states are protected by this symmetry~\cite{furusaki.17}.

\begin{figure*}
\includegraphics[width=\linewidth]{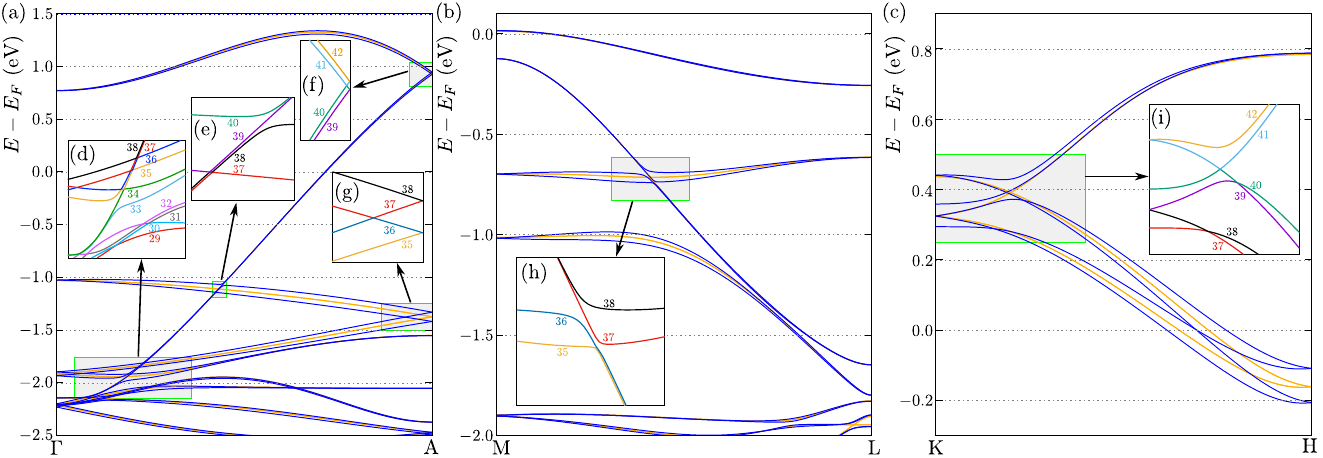}
\caption{
Effect of SOC on the electronic band structure close to the Fermi level along $\Gamma$--A (a), M--L (b), and K--H (c).
Results without and with SOC are shown by orange and blue lines, respectively.
Several zoomed-in views are presented as insets (d)--(i) to highlight the crossing and avoided-crossing behavior.
The band indices are also indicated.
\label{fig.el_soc}}
\end{figure*}

\begin{figure*}
\includegraphics[width=\linewidth]{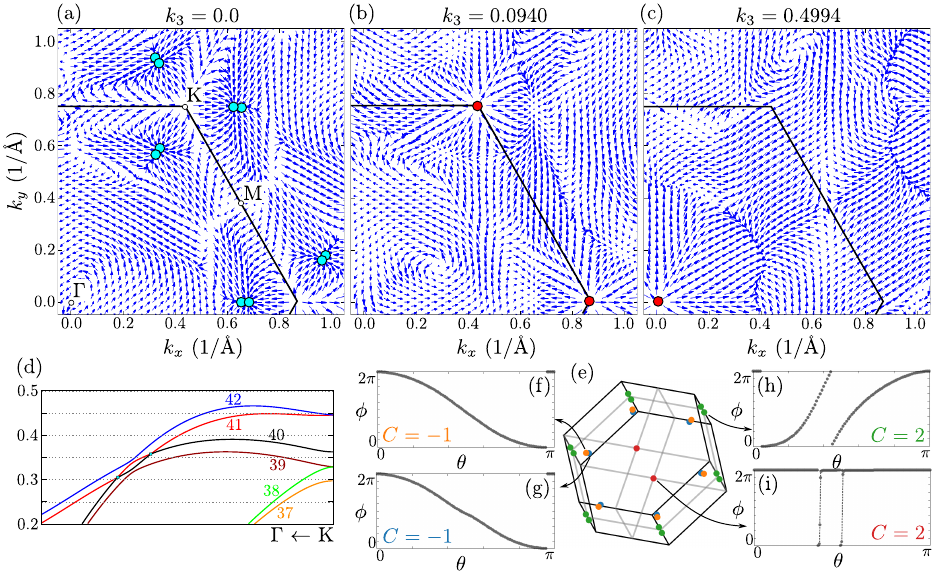}
\caption{
Topological properties of the nodal points between the 40th and 41st bands.
(a)--(c) Berry curvature field in the $k_{xy}$ planes at different $k_{3}$ values (as labeled), intersecting the Weyl points.
Red and cyan points indicate sources and sinks of the Berry curvature, respectively.
(d) Electronic band structure along the $\Gamma$--K path, showing the nodal points (for the $\Gamma$--A and K--H paths, see Fig.~\ref{fig.el_soc}).
(e) Locations of the Weyl points in the BZ.
(f)--(i) Corresponding evolution of the sum of the Wannier charge centers on spheres enclosing the Weyl points.
\label{fig.el_node}}
\end{figure*}

In the absence of SOC, fourfold band degeneracies are observed at the high-symmetry points $\Gamma$, A, K, and H.
In the presence of $\mathcal{T}$, such points arise from crossings between two doubly degenerate bands.
When SOC is included, these band degeneracies are lifted, reducing the fourfold degeneracies to twofold degeneracies.
However, we should note that the $\Gamma$, A, M, and L points are time-reversal-invariant momenta (TRIM) of the $P6_{2}$22 ($P6_{4}$22) SG.
By definition, all bands at TRIM remain doubly degenerate in the presence of SOC.
Beyond the TRIM, we can also find some doubly degenerate points at K and H, even in the presence of SOC.
This behavior is presented in Fig.~\ref{fig.el_soc}, which shows the bands close to the Fermi level along the $\Gamma$--A, M--L, and K--H paths.

Additionally, along the $\Gamma$--A, M--L, and K--H paths, as well as throughout the BZ, the SOC-induced splitting of bands can give rise to new crossing or avoided-crossing points.
This behavior is clearly visible in Figs.~\ref{fig.el_soc}(e)--\ref{fig.el_soc}(i).
For example, new crossings, along the $\Gamma$--A direction, the initially doubly degenerate bands (orange lines) split into two predominantly parallel branches (blue lines), as shown in Fig.~\ref{fig.el_soc}(g).
This behavior can be observed at several locations in the band structure, with varying strengths [cf. Fig.~\ref{fig.el_soc}(f)].
An avoided crossing between decoupled bands along M--L is presented in Fig.~\ref{fig.el_soc}(h).
Here, all bands split by SOC exhibit avoided crossings.
A more complex situation can be observed along the $\Gamma$--A and K--H paths.
The introduction of SOC leads to a complex band structure containing both crossing and avoided-crossing bands [see Figs.~\ref{fig.el_soc}(d),~\ref{fig.el_soc}(e), and~\ref{fig.el_soc}(i)].

As mentioned above, the nonsymmorphic chiral symmetries play an important role in this structure.
In fact, the $\Gamma$--A path is invariant under both the $C_{3z}$ and $C_{6z}$ screw rotation symmetries, while K--H is invariant only under $C_{3z}$.
For all ${\bm k}$ along these paths, the system Hamiltonian $H({\bm k})$ commutes with the screw rotation operator $C_{nz}$ ($n = 3$ or $6$), and the energy eigenstates are also eigenstates of $C_{nz}$~\cite{tsirkin.souza.17}.
In practice, all bands along $\Gamma$--A and K--H can also be labeled by a {\it rotation label} $m$.
In the absence of SOC, $m$ takes integer values (from $0$ to $n-1$), while in the presence of SOC, $m$ takes half-integer values [from $1/2$ to $(2n-1)/2$] due to the she SU(2) symmetry.
This feature leads to the realization of a three-band structure along K--H, with nodes at K and H [see Fig.~\ref{fig.el_soc}(c)].
A similar structure, consisting of six bands, is found along $\Gamma$--A.
Finally, we should note that band crossings along $\Gamma$--A and K--H can occur only between bands with different values of $m$.
For bands with the same $m$, avoided crossings occur.
This rule leads to the structures presented in Figs.~\ref{fig.el_soc}(d) and \ref{fig.el_soc}(i), where both crossing and avoided-crossing points can be observed.

Let us now analyze the topological features of some of the band crossings.
Beyond the topological properties related to the nonsymmorphic chiral symmetry discussed above, SOC can also induce additional nodal points.
To investigate the topological properties of the Weyl points, we focus on the touching points between the 40th and 41st bands, which, as indicated in Figs.~\ref{fig.el_soc}(f) and~\ref{fig.el_soc}(i), exhibit interesting crossings around the A and K points.
The obtained results are presented in Fig.~\ref{fig.el_node}.
First, we calculate the Berry curvature field in the $k_{xy}$ planes containing these interesting points (at different $k_{3}$ values, as indicated).
The Berry curvature is given by
\begin{eqnarray*}
\Omega ( {\bm k} ) = {\bm \nabla}_{\bm k} \times A ( {\bm k} ) ,
\end{eqnarray*}
where
\begin{eqnarray*}
A ( {\bm k} ) = \sum_{n=1}^\text{N} i \langle u_{n{\bm k}} \vert {\bm \nabla}_{\bm k} \vert u_{n{\bm k}} \rangle.
\end{eqnarray*}
As can be seen in Fig.~\ref{fig.el_node}(b), at $k_{3} = 0.0940$, a source of Berry curvature is clearly visible at the corner of the BZ contour (black line).
This point corresponds to the touching point at $(1/3,1/3,0.0940)$ shown in Fig.~\ref{fig.el_soc}(i).
A similar situation is found in the $k_{3} = 0.4994$ plane, where the SOC-induced touching point shown in Fig.~\ref{fig.el_soc}(d) is located.
In this case, the point at $(0,0,0.4994)$, i.e., close to the A point, acts as a source of Berry curvature.
Additionally, in the $k_{3} = 0$ plane, we find two Weyl points along the $\Gamma$--K direction, which act as sinks of Berry curvature [Fig.~\ref{fig.el_node}(a)].
The presence of this pair of Weyl points is also clearly visible in the band structure along the $\Gamma$--K path [see Fig.~\ref{fig.el_node}(d)].
The positions of all Weyl points between the 40th and 41st bands in the BZ are presented in Fig.~\ref{fig.el_node}(e).
For the identified Weyl points, the Chern numbers $C$ can be obtained from the evolution of the sum of the Wannier charge centers on a sphere enclosing each Weyl point [see Figs.~\ref{fig.el_node}(f)--\ref{fig.el_node}(i)].
We find that the Weyl points in the $k_{3} = 0$ plane, at energies of $0.3038$ and $0.3575$~eV, have $C = -1$, while the other Weyl points have $C = 2$ (at energies of $0.3956$ and $0.9375$~eV).
In summary, for the Weyl nodes between the 40th and 41st bands, we find six pairs of Weyl nodes with $C = -1$ [orange and blue points on Fig.~\ref{fig.el_node}(e)] and six Weyl nodes with $C = 2$ [red and green points on Fig.~\ref{fig.el_node}(e)].
Moreover, the total topological charge of all Weyl points is zero, consistent with the Nielsen--Ninomiya no-go theorem~\cite{nielsen.ninomiya.81,nielsen.ninomiya.81b}.

Due to the relatively small magnitude of SOC, the investigation of Weyl points is quite challenging.
In practice, we identify more than 10,000 touching points, some of which can be recognized as Weyl nodes -- some of them close to the Fermi level are collected in Table~\ref{tab.weyl}.
However, an accurate determination of the Weyl-point chirality $C$, based on the evolution of the sum of the Wannier charge centers on a sphere enclosing the Weyl point, requires higher numerical precision.
This is related to the rapid evolution of the sum of the Wannier charge centers, particularly for the SOC-induced Weyl nodes [as shown in Fig.~\ref{fig.el_node}(i)].

As we mentioned earlier, the absence of $\mathcal{I}$ symmetry in the $P6_{2}22$ (or $P6_{4}22$) SG allows Weyl points to emerge.
It should be noted that changing the SG from $P6_{2}22$ to $P6_{4}22$, and {\it vice versa}, does not alter the main features of the electronic band structure. 
However, this transformation reverses the chirality of the observed surface states, corresponding to a simultaneous reversal of the sign of the Weyl-point topological charge, i.e., the Chern number.

\begin{figure}
\includegraphics[width=\linewidth]{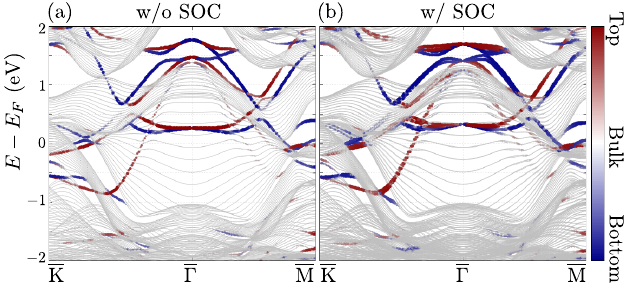}
\caption{
The electronic band structure in the slab geometry in the absence (a) and presence (b) of the SOC.
The color and size of the dots indicate the contribution from surface atoms.
Results are shown for a slab with a (001) surface.
\label{fig.slab}}
\end{figure}

\begin{figure}
\includegraphics[width=\linewidth]{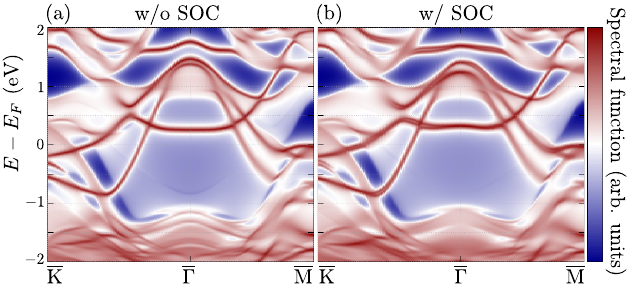}
\caption{
Surface spectral function in the (a) absence and (b) presence of SOC. 
Results for the (001) surface, corresponding to the top surface of the slab shown in Fig.~\ref{fig.slab}.
\label{fig.ss}}
\end{figure}

\begin{figure*}
\includegraphics[width=\linewidth]{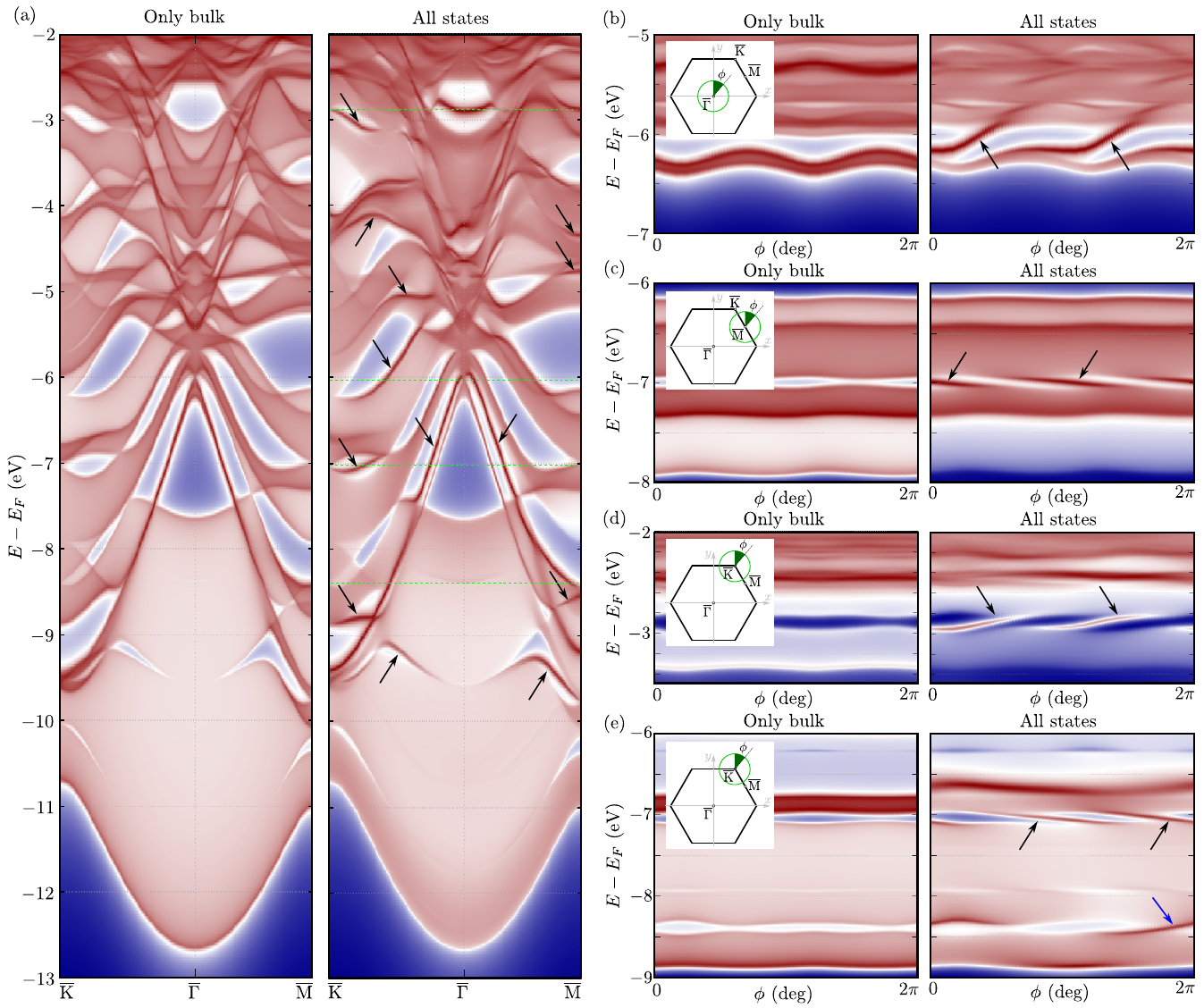}
\caption{(a) Spectral function along the $\overline{\text{K}}$--$\overline{\Gamma}$--$\overline{\text{M}}$ path for the bulk and all states (as labeled).
Black arrows indicate selected surface states. 
The green dashed line marks the binding energies corresponding to the constant-energy contours presented in Fig.~\ref{fig.arcs}. 
Spectral functions along closed loops surrounding the $\overline{\Gamma}$ (b), $\overline{\text{M}}$ (c), and $\overline{\text{K}}$ (d,e) points, considering the bulk and all states, as indicated. Each loop is chosen as a circle centered at the corresponding high-symmetry point, with a radius of $0.05$~\AA$^{-1}$, as schematically illustrated in the inset (not to scale). 
Clearly visible chiral states are highlighted by black arrows. 
All results are obtained in the presence of SOC.
The color scale is the same as in Fig.~\ref{fig.ss}.
\label{fig.ss_chiral}}
\end{figure*}

\begin{figure*}
\includegraphics[width=0.7\linewidth]{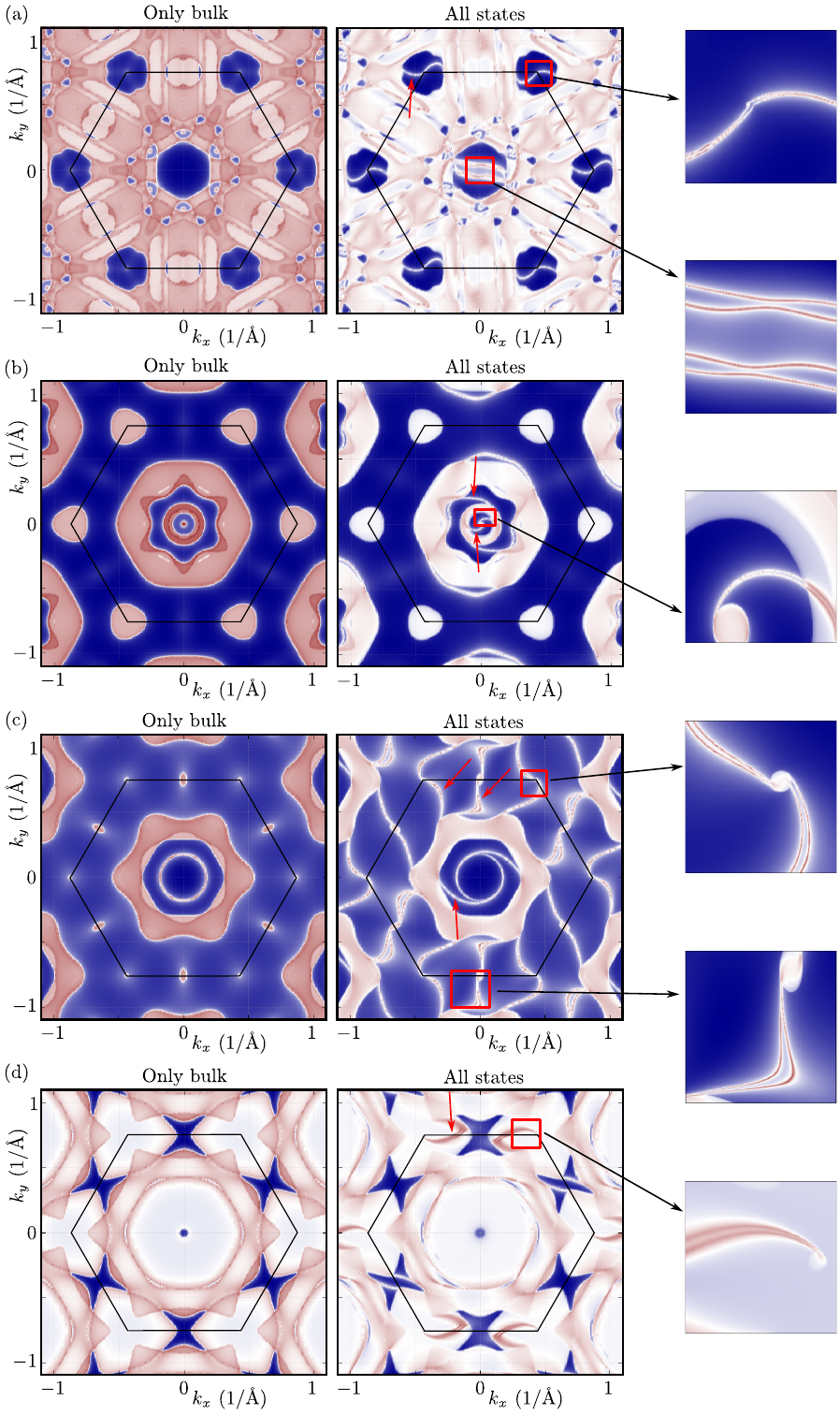}
\caption{
Constant-energy contours at selected binding energies, intersecting the chiral edge modes indicated in Fig.~\ref{fig.ss_chiral}: 
(a) $-2.88$~eV (chiral edge mode around $\overline{\text{K}}$), 
(b) $-6.08$~eV (chiral edge mode around $\overline{\Gamma}$), 
(c) $-7.02$~eV (chiral edge modes around $\overline{\text{M}}$ and $\overline{\text{K}}$), and (d) $-8.39$~eV (chiral edge mode around $\overline{\text{K}}$). 
Red arrows indicate the locations of the chiral edge modes. 
The smaller panels on the right show high-resolution views of the regions highlighted by red squares. 
Results are obtained in the presence of SOC, considering both the bulk and all states, as indicated. The color scale is the same as in Fig.~\ref{fig.ss}.
\label{fig.arcs}}
\end{figure*}

\begin{table*}[!t]
\caption{
\label{tab.weyl}
Parameters of selected Weyl nodes near the Fermi level in the presence of SOC, located at high-symmetry points. 
Only one representative of each symmetry-equivalent set of $\bm{k}$-points is listed, together with its multiplicity in the BZ.
}
\centering
\begin{ruledtabular}
\begin{tabular}{cccccccc}
\textbf{Bands} & \textbf{Energy} & \multicolumn{3}{c}{${\bm k}$ (fractional)} & $\mathcal{C}$ & \textbf{Multiplicity} & \\
& (eV) & $k_{1}$ & $k_{2}$ & $k_{3}$ & & & \\
\hline
%%%%%%%%%%%%%%%%%%%%%%%%%%%%%
\multicolumn{8}{c}{Several Weyl points around the Fermi level} \\
\hline
%%%%%%%%%%%%%%%%%%%%%%%%%%%%%
41-42 & 1.39131 & 1/2 & 1/2 & 1/2 & -1 & 3 & L  \\
41-42 & 1.03911 & 1/2 & 0 & 0.23329 & 1 & 6 & M-L   \\
41-42 & 0.94728 & 0 & 0 & 1/2 & 1 & 1 & A  \\
39-40 & 0.92794 & 0 & 0 & 1/2 & 1 & 1 & A  \\
41-42 & 0.78918 & 1/3 & 1/3 & 1/2 & -1 & 2 & H  \\
41-42 & 0.77234 & 0 & 0 & 0 & -1 & 1 & $\Gamma$  \\
41-42 & 0.77223 & 1/2 & 0 & 0.0457 & -1 & 6 & M-L   \\
41-42 & 0.76392 & 1/3 & 1/3 & 0.37324 & 1 & 4 & K-H  \\
41-42 & 0.76049 & 1/2 & 0 & 0 & 1 & 3 & M  \\
41-42 & 0.44526 & 1/3 & 1/3 & 0 & 1 & 2 & K  \\
40-41 & 0.39562 & 1/3 & 1/3 & 0.09403 & 1 & 4 & K-H  \\
39-40 & 0.36826 & 1/3 & 1/3 & 0.12161 & 1 & 4 & K-H  \\
38-39 & 0.32807 & 1/3 & 1/3 & 0 & -2 & 2 & K  \\
37-38 & 0.29439 & 1/3 & 1/3 & 0.07298 & -1 & 4 & K-H  \\
39-40 & 0.02052 & 0 & 1/2 & 0 & -1 & 3 & M  \\
39-40 & -0.05184 & 1/2 & 1/2 & 0.15744 & 1 & 6 & M-L   \\
38-39 & -0.11988 & 1/3 & 1/3 & 0.43508 & -1 & 4 & K-H   \\
37-38 & -0.12046 & 0 & 1/2 & 0 & 1 & 3 & M  \\
39-40 & -0.15859 & 1/2 & 0 & 0.28399 & -1 & 6 & M-L  \\
37-38 & -0.20952 & 0 & 1/2 & 0.06899 & 1 & 6 & M-L  \\
39-40 & -0.22816 & 1/2 & 0 & 0.39299 & 1 & 6 & M-L  \\
39-40 & -0.25200 & 0 & 1/2 & 1/2 & -1 & 3 & L  \\
37-38 & -0.55545 & 0 & 1/2 & 0.16796 & -1 & 6 & M-L  \\
37-38 & -0.61261 & 1/2 & 0 & 1/2 & 1 & 3 & L  \\
35-36 & -0.69609 & 1/2 & 1/2 & 0 & -1 & 3 & M  \\
33-34 & -0.73924 & 1/3 & 1/3 & 1/2 & 2 & 2 & H  \\
34-35 & -0.74068 & 1/3 & 1/3 & 0.45317 & 1 & 4 & K-H  \\
35-36 & -0.75877 & 0 & 1/2 & 0.21373 & -1 & 6 & M-L  \\
35-36 & -0.91429 & 1/3 & 1/3 & 0.34905 & -2 & 4 & K-H  \\
34-35 & -0.93133 & 1/3 & 1/3 & 0.3419 & -1 & 4 & K-H  \\
33-34 & -1.01447 & 0 & 1/2 & 0 & 1 & 3 & M  \\
39-40 & -1.02129 & 0 & 0 & 0 & 1 & 1 & $\Gamma$  \\
39-40 & -1.06333 & 0 & 0 & 0.14794 & -1 & 2 & $\Gamma$-A  \\
35-36 & -1.30487 & 1/2 & 1/2 & 0.35214 & -1 & 6 & M-L  \\
35-36 & -1.41442 & 0 & 0 & 1/2 & 1 & 1 & A  \\
35-36 & -1.49092 & 1/3 & 1/3 & 0.06625 & -2 & 4 & K-H  \\
%%%%%%%%%%%%%%%%%%%%%%%%%%%%%
\end{tabular}
\end{ruledtabular}
\end{table*}

\section{Surface properties}
\label{sec.surf}

In this section, we discuss the electronic states emerging at the surface of the system. The bulk topological properties discussed in the previous section are manifested in the surface states, which arise from the breaking of translational symmetry along the surface-normal direction. 
We focus on the surface states associated with the chiral Si chains. 
Therefore, we consider the (001) surface, which is perpendicular to the direction of the chains.

We begin by discussing the electronic band structure in the slab-like geometry for the (001) surface, shown in Fig.~\ref{fig.slab}.
As in the bulk band structure, the inclusion of the SOC lifts the band degeneracy.
Nevertheless, the main features associated with the bulk states remain unchanged (gray lines).
The surface states (blue and red dots) exist independently of the SOC [cf.~the left and right panels].
Here, the SOC also leads to a splitting of the surface states, which is clearly visible, for example, around the $\Gamma$ point at $1.3$~eV.

Using the obtained tight-binding model, we also calculate the spectral Green's function for the (001) surface, both in the absence and presence of SOC [Figs.~\ref{fig.ss}(a) and~\ref{fig.ss}(b), respectively].
Near the Fermi level, at approximately $0.5$~eV around the $\Gamma$ point, a nearly dispersionless surface state is observed. As mentioned earlier, the inclusion of SOC leads to a splitting of the surface states. Nevertheless, this splitting is relatively small, and the main features of the spectral function remain essentially unchanged.

Note that the results presented in Figs.~\ref{fig.slab} and~\ref{fig.ss} were obtained in the absence and presence of SOC, respectively. 
Upon inclusion of SOC, the surface states can split, as previously reported in the context of chiral cubic compounds, such as PtGa~\cite{yao.manna.20} and EuPtSi~\cite{mahraj.ptok.25}. 
This splitting originates from SOC-induced lifting of the degeneracy between the surface states. Nevertheless, the surface states themselves persist independently of SOC.

The electronic band structure around the Fermi level is rather complex, particularly in the energy range where the Nb $d$ and Si $p$ are strongly hybridized (i.e.,~around $-2$~eV).
As can be seen, the surface states are predominantly located within the energy range occupied by the bulk bands. 
Consequently, these surface states may be difficult to identify experimentally.
Indeed, the experimental ARPES spectra~\cite{zhang.shishidou.23} are complex and difficult to interpret.
Moreover, direct observation of the surface states using ARPES measurements has not yet been reported.

Much clearer results are obtained at energies deeply well below the Fermi level ($\ll - 2$~eV), where the Nb $d$ states play a less significant role.
The spectral function for these deeper-lying states is presented in Fig.~\ref{fig.ss_chiral}.
The spectral function along the $\overline{\text{K}}$--$\overline{\Gamma}$--$\overline{\text{M}}$ path was calculated [Fig.~\ref{fig.ss_chiral}(a)], as well as along closed loops around the $\overline{\Gamma}$, $\overline{\text{M}}$, and $\overline{\text{K}}$ points [Figs.~\ref{fig.ss_chiral}(b)--\ref{fig.ss_chiral}(e)].
For better readability, we present the spectral function calculated only for the bulk states (left panels) and for all states, including the surface states (right panels).
As can be seen, the surface states are present throughout the entire energy range (marked by black arrows).
Some of these surface states appear between the Weyl points, which are visible in the spectral function as cone-like structures.

\subsection{Chiral edge modes}

First, we would like to point out that chiral symmetry guarantees the existence of chiral edge modes.
For such modes, the surface states propagate in a single direction along a finite closed loop. 
To demonstrate this feature in NbSi$_2$, we consider a circle with a radius of $0.05$~\AA$^{-1}$ centered at the $\overline{\Gamma}$, $\overline{\text{M}}$, and $\overline{\text{K}}$ points [see Figs.~\ref{fig.ss_chiral}(b)--\ref{fig.ss_chiral}(e)].
The chiral edge modes (marked by black arrows) are clearly visible within a certain energy range. 
In constant-energy contours, the chiral edge modes can be readily identified as single ``lines'' connecting different groups of bulk states.

The number of propagating chiral modes corresponds to the absolute value of the effective Chern number $|C|$.
Indeed, a closed loop encloses all Weyl points within its area.
In the absence of SOC, the chiral mode around $-8.39$~eV can be associated with a Weyl point with an effective $|C| = 1$, corresponding to one propagating mode [marked by the red arrow in Fig.~\ref{fig.ss_chiral}(e)].
In contrast, the other modes [marked by black arrows in Figs.~\ref{fig.ss_chiral}(b)--\ref{fig.ss_chiral}(d)] arise from the presence of Weyl nodes with an effective $|C| = 2$.
Please note that the sign of $C$ determines the propagation direction of the chiral mode [cf. the modes in Figs.~\ref{fig.ss_chiral}(c) and~\ref{fig.ss_chiral}(d)].
As mentioned earlier, in the presence of SOC, the surface states become decoupled, and a more detailed analysis is required.

\begin{figure*}
\includegraphics[width=0.75\linewidth]{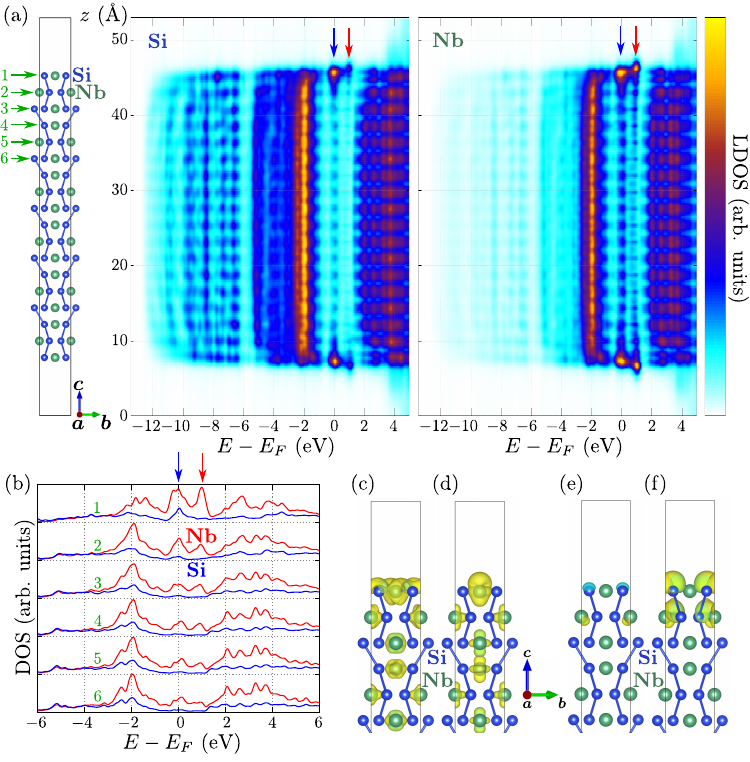}
\caption{
DFT-based investigation of the real-space localization of the edge modes in the presence of SOC. 
(a) Spatially resolved DOS for Si and Nb atoms, as indicated. 
The slab is viewed from the left side. (b) Site-resolved DOS for Si and Nb atoms, as indicated. The green numbers in panel (a) indicate the site indices, counted from the top surface of the slab. 
Red and blue arrows indicate the positions of the peaks in the DOS. 
Partial charge densities calculated by considering all states within the energy ranges of $-0.5$--$0.5$~eV (c) and $0.5$--$1.25$~eV (d), as well as two individual states at the $\overline{\Gamma}$ point corresponding to the edge modes (e,f). 
The isosurfaces correspond to charge densities of $5\times10^{-3}$~e/\AA$^{-3}$ for panels (c,d) and $1\times10^{-3}$~e/\AA$^{-3}$ for panels (e,f).
\label{fig.ss_dos}}
\end{figure*}

For further analysis, we select energies at which the chiral edge modes are well distinguished from the rest of the spectrum [regions marked by black arrows in Figs.~\ref{fig.ss_chiral}(b)--\ref{fig.ss_chiral}(e)].
The energies $-2.88$, $-6.08$, $-7.02$, and $-8.39$~eV are marked by horizontal green dashed lines in Fig.~\ref{fig.ss_chiral}(a), while the corresponding constant-energy contours are presented in Fig.~\ref{fig.arcs}.
First, for the selected energies, we can readily identify the positions of Weyl nodes at high-symmetry points.
These Weyl points originate from the projection of the bulk Weyl points onto the two-dimensional (slab) BZ.
As can be seen, the actual number of Fermi arcs (chiral modes) connecting to these points is doubled in the presence of SOC.
This is clearly visible in the enlarged regions of the BZ shown in the small panels on the right-hand side of Fig.~\ref{fig.arcs}.
In each case, a high-resolution scan of the constant-energy contours reveals a doubled surface-state structure.
For example, we observe four Fermi arcs connected to the $\overline{\text{K}}$ point in Figs.~\ref{fig.ss_chiral}(a) and~\ref{fig.ss_chiral}(c).
The same behavior is observed for the Fermi arcs connected to $\overline{\Gamma}$ in Fig.~\ref{fig.ss_chiral}(b).
This may suggest either (i) the existence of a single Weyl point with an effective $|C| = 4$ or (ii) the existence of two Weyl points with an effective $|C| = 2$ at nearby energies.
Additionally, chiral modes are also visible in the center of the BZ in Figs.~\ref{fig.ss_chiral}(b) and~\ref{fig.ss_chiral}(c).
In this case, the surface states connect the inner bulk states to the outer bulk states.

\subsection{Real space localization}

Several techniques can be used to investigate the real-space localization of the observed states.
In Fig.~\ref{fig.ss_dos}, we present the spatially resolved DOS~\cite{lodeiro.rauch.22}, site-dependent DOS, and partial charge density.
The results were obtained for a slab geometry consisting of 54 atoms (18 formula units), corresponding to a slab thickness of approximately $37.5$~\AA.
Due to the symmetry of the system, the top and bottom (001) surfaces have the same termination [see Fig.~\ref{fig.ss_dos}(a)].

The spatially resolved DOS is presented over a wide energy range [see Fig.~\ref{fig.ss_dos}(a)] and is divided into contributions from Si and Nb atoms.
Throughout the entire energy range, the bulk states form a standing-wave-like structure, with a mostly homogeneous distribution of the local DOS (LDOS).
The local maxima of the LDOS correspond to the exact positions of the atoms in the slab (cf. the LDOS and the crystal structure on the left).
As mentioned earlier, the deep states are mostly associated with the Si atoms (p orbitals), while around $-2$~eV, we observe a strong mixing of states throughout the entire system.
For the deep states, the edge modes can be difficult to identify due to their small contribution to the total LDOS.
Nevertheless, two pronounced peaks are clearly visible around the Fermi level (marked by blue and red arrows).
These peaks are visible in both the Si and Nb LDOSs.
In the case of Si, the DOS decreases significantly, whereas for Nb, a relatively strong contribution from bulk states remains in the central part of the slab.
This behavior is also visible in the site-dependent DOS presented in Fig.~\ref{fig.ss_dos}(b).
Moving from the surface towards the central part of the slab (from the 1st to the 6th layer), the observed peaks around $0$ and $1$~eV decrease.
However, a significant difference between the behavior of Si and Nb is observed.
The DOS for Nb decreases by only a factor of two, whereas for Si, it decreases by approximately a factor of ten.
This effect can be related to the different contributions of Si and Nb states to the bulk DOS, which is mostly dominated by Nb states.
Nevertheless, we should remember that a significant decrease in the LDOS is expected for edge modes, which should contribute predominantly at the surface.
This may suggest an important role of the Si chains in the localization of the edge modes.

Finally, we calculate the partial charge density in the energy range around the previously reported peaks [Figs.~\ref{fig.ss_dos}(c)--\ref{fig.ss_dos}(f)].
The partial charge densities are calculated in two ways:
(i) within selected energy ranges containing the previously discussed peaks, and
(ii) for individual states at the $\overline{\Gamma}$ point corresponding to the edge modes.
In the first case, the partial charge density captures contributions from both edge and bulk states.
We select the energy ranges $-0.5$--$0.5$~eV and $0.5$--$1.25$~eV [Figs.~\ref{fig.ss_dos}(c) and~\ref{fig.ss_dos}(d), respectively].
Here, we present iso-surfaces corresponding to $5 \times 10^{-3}$~e/\AA$^{3}$.
We observe contributions from bulk states throughout the entire system, while an enhanced charge density on Nb atoms is visible near the surfaces.
A different behavior is observed in the second case, where the partial charge density is calculated for individual states.
As discussed earlier, the edge modes at $\overline{\Gamma}$ are degenerate.
Thus, we obtain four states (doubled due to the contributions from the top and bottom surfaces) with symmetric partial charge density distributions.
Two distinguishable distributions are presented in Figs.~\ref{fig.ss_dos}(e) and~\ref{fig.ss_dos}(f).
As can be seen, in this case, the charge density is mostly localized near the surfaces, around the Si atoms.

All results obtained from the partial charge density are consistent with the spatially resolved and site-dependent DOS.
The strong contribution of Nb states to the DOS indicates a charge distribution throughout the entire system, while the accumulation of Si DOS near the surfaces suggests charge localization around the outermost Si atoms.

%%%%%%%%%%%%%%%%%%%%%
%%%%%%%%%%%%%%%%%%%%%
%%%%%%%%%%%%%%%%%%%%%

\section{Summary}
\label{sec.sum}

In this paper, we investigate the electronic band structure of nonsymmorphic chiral NbSi$_2$ with hexagonal symmetry.
We find that the SOC is relatively weak, and leads to band splittings of approximately $20$--$27$~meV for bands close to the Fermi level, in agreement with experimentally reported values. 
Owing to its nonsymmorphic chiral crystal structure, the system hosts topological surface states, including chiral edge modes. 
The observed surface states persist in the absence of SOC, demonstrating that the topological properties originate from the nonsymmorphic chiral symmetry rather than from SOC.

The absence of inversion symmetry allows Weyl nodes to emerge, while the chiral symmetry gives rise to chiral edge modes.
We investigate the bulk topological properties through the Berry curvature, demonstrating the existence of Weyl nodes.
We find that some Weyl points carry a topological Chern number of $C = \pm 2$.
The corresponding chiral modes around high-symmetry points are also demonstrated through an analysis of the surface spectral function.

Some of the surface states occur in the immediate vicinity of the Fermi level and can also be observed in the DOS as well-localized states.
We investigate this behavior using spatially resolved and site-resolved DOS.
The peaks observed in the DOS near the slab surfaces indicate the presence of well-localized surface states.
We observe an enhancement of the DOS at Nb atoms and charge localization around Si atoms near the surfaces.

In summary, we demonstrate that NbSi$_{2}$ is a topological system.
The absence of magnetism and the presence of nonsymmorphic chiral symmetry make this system an excellent platform for the experimental investigation of its topological properties.
The existence of well-localized edge modes around the Fermi level opens up the possibility of high-resolution STM studies of the surface states.

\begin{acknowledgments}
Some figures in this work were rendered using {\sc Vesta}~\cite{momma.izumi.11} software.
This work was supported by the National Science Centre (NCN, Poland) under Project No.~2025/59/B/ST3/00232.
\end{acknowledgments}

%%%%%%%%%%%%%%%%%%%%%
%%%%%%%%%%%%%%%%%%%%%
%%%%%%%%%%%%%%%%%%%%%

%\nocite{*}
\bibliography{biblio.bib}

\end{document}